\documentstyle[a4wide,12pt]{article}
\begin{document}
\begin{flushright}
\baselineskip=12pt
{SUSX-TH-95/15}\\
July 1995
\end{flushright}

\begin{center}
{\huge\bf The $b{\rightarrow}s{\gamma}$ constraint in
effective supergravities from  string theory\\}
\vglue 0.4cm
{G. V. KRANIOTIS \footnote{e.mail address:G.Kraniotis@sussex.ac.uk
}\\}
{\it School of Mathematical and Physical Sciences \\}
{\it University of Sussex,\\}
{\it Brighton BN1 9QH, United Kingdom \\}
\baselineskip=12pt

\vglue 0.25cm
ABSTRACT
\end{center}
{\rightskip=3pc
\leftskip=3pc
\noindent
\baselineskip=20pt
We study the constraints from the $b{\rightarrow}s{\gamma}$ decay
in the parameter space of effective supergravities from orbifold
string theory and with minimal
supesymmetric particle content. Both the
general dilaton-dominated universal scenario as well as
a non-universal scenario for the soft terms are
investigated. It is found that the recently reported CLEO upper
and lower bounds constrain the parameter space of the models under
scrutiny. In particular we find constraints on the
values of the parameter $\tan{\beta}$ and the gluino
masses. In this class of string scenarios the negative sign of the
Higgs mixing parameter $\mu$, is phenomenologically preferred.}

\vfill\eject
\setcounter{page}{1}
\pagestyle{plain}
\baselineskip=14pt
\section {Introduction}
\vglue 0.1cm
One of the prime tasks of the Large Hadron Collider is to search
for the supersymmetric partners of the Standard Model multiplets.
Once the first sparticles are discovered, the program of
sparticle spectroscopy will give us vital clues for the {\em underlying}
theory that explains the observed spectrum. As  is well known
$N=1$ superymmetric field theories predict a very rich structure
of sparticles from a few fundamental parameters. These parameters
break the supersymmetry softly at an energy of the order of the
electroweak scale thus ensuring that the hierarchy problem is at
least technically solved and that the superparticles do not have the
same mass as their Standard Model partners. It is customary to
parametrize the effects of the soft-supersymmetry breaking terms
by four universal parameters: the universal gaugino mass $M_{1/2}$,
 the scalar terms associated with the trilinear
couplings in the superpotential $A$, the scalar masses $m_{0}$,
and the $B$ term
associated with the Higgs-doublet mixing term in the superpotential
\footnote {Note however, that this parametrization is not the most
general case since in string theory
non-universality of the soft terms is very common \cite{Iba:Spain}.}.

On the other hand,
in order to be able to interpret the ``lines'' in the sparticle
spectrum we need to have a theory which will differentiate among
the many alternatives of the soft-parameter space.
The only example of the sort of theory we are aiming for is heterotic
string theory. In string theory these soft susy- breaking parameters are in
principle, calculable, but a  definite answer
is at present lacking due to the fact that the superymmetry breaking
mechanism in the theory is not well understood. However,
in the pioneering work of \cite{Iba:Spain}  the effect
of SUSY-breaking is parametrized by the VEVs of the $F$-terms of the dilaton
$(S)$ and the moduli $(T_m)$ chiral superfields, generically present
in large classes of four-dimensional supersymmetric heterotic strings.
This is an important step towards a theory which will explain the rich
sparticle spectrum. In this work the soft-parameter space has been
reduced since many interesting relations among the soft-parameters
have been found which in principle can be tested at the LHC .

	However, until the first experiments at LHC start running,
we have to use all the current experimental information in order to
study the parameter space of the effective supergravities from
string theory. Unfortunately, most of the precision LEP measurements
are not very sensitive to new physics as the Standard Model
contributions enter at the tree level, while possible new physics
contributions begin at the one loop level. Thus the most one might hope
for in these measurements is a few percent correction from new
physics.

However, it has become well known that the $b{\rightarrow}s{\gamma}$
decay is an exception to this and that is
a powerful tool for testing Physics beyond the Standard
Model \cite{GUIDICE,Bert:Nucl}. This is  the case for several reasons:
First as a FCNC process, the $b{\rightarrow}s{\gamma}$ decay, arises
first at the one loop level so that the Standard Model loops and new
physics loops enter at the same level. Second, the decay is
of size $G^{2}_{F}{\alpha}$, where $G_{F}$ is the Fermi constant
(rather then $G_{F}^{2}{\alpha}^{2}$ as is usual for FCNC
processes). Third, there are available experimental data on the
exclusive \cite{CLEO:Ammar} $B{\rightarrow}K^{\ast}{\gamma}$ and
the inclusive \cite{Brow:Bdec} $B{\rightarrow}X_{s}{\gamma}$
decays that lead to upper and lower bounds on the branching
ratio BR($b{\rightarrow}s{\gamma}$) of the same order as
the SM prediction. In particular,
from the inclusive $B$ decay:$1{\times}10^{-4}$$<$BR($b{\rightarrow}
s{\gamma})$$<$$4{\times}10^{-4}$ \cite{Glasgow}. In the Minimal Supersymmetric
extension of the Standard Model there are additional contributions
to the decay besides the SM diagram with a $W$ gauge boson and
a top quark in the loop. In particular, there are additional
contributions coming from loops involving charged Higgses $(H^{-})$
and a top quark, charginos $(\chi^{-})$ and u-type squarks
(of which the relevant contributions come from the stops,
${\tilde{t_{L,R}}}$, and scharms, ${\tilde{c}}$, and a gluino
or neutralinos $({\chi_{i}^{0}})$ plus a d-type squark (mainly
${\tilde{b}}$ and ${\tilde{s}})$ \cite{Bert:Nucl}. As pointed
in ref. \cite{Bert:Nucl}, the latter two diagrams do not
contribute significantly to the $BR$ and can therefore be
neglected. It is  common practice to use the ratio defined as
\begin{equation}
R=\frac{BR(b{\rightarrow}s{\gamma})}{BR(b{\rightarrow}ce{\bar{\nu}})}
{\simeq}\frac{BR(B{\rightarrow}X_{s}{\gamma})}{BR(B{\rightarrow}X_{c}
e{\bar{\nu_{e}}})}
\label{Rat:1}
\end{equation}
to constrain various models, utilizing the well determined value of
$ 10.7\pm0.5\%$ for $BR(B{\rightarrow}X_{c}e{\bar{\nu_{e}}})$. The
advantage of using {\it R},
instead of $BR(b{\rightarrow}s{\gamma})$, is
that the latter is dependent
upon $m^{5}_{b}$ while the former
only depends on $z=m_{c}/m_{b}$, the ratio between the {\it c} and
{\it b}
quark masses, which is much better determined than both masses,
i.e., $z=0.316\pm0.013$ \cite{Max:Plan}.

The ratio R defined in Eq. $(\ref{Rat:1})$ is given by \cite{Bert:Nucl}
\begin{equation}
R=\frac{{|V_{ts}^{\ast}V_{tb}|}^{2}}{{|V_{cb}|}^{2}}\frac{6{\alpha_{QED}}}
{\pi}\frac{{{[\eta^{16/23}}A_{\gamma}+\frac{8}{3}({\eta^{14/23}}-
{\eta^{16/23}})A_{g}+C]}^{2}}{I(m_{c}/m_{b})[1-(2/3{\pi}){\alpha_{s}(m_{b})}
f(m_{c}/m_{b})]}
\label{Rat:2}
\end{equation}
where $\eta=\alpha_{s}(M_{W})/\alpha_{s}(m_{b})$, and $M_{W}$
is the $W$ boson mass. Here, $I(z)=1-8z^{2}+8z^{6}-z^{8}-24z^{4}\ln{z}$
is the phase-space factor, and $f(z)=2.41$, is a QCD correction factor,
for the semileptonic process, $b{\rightarrow}ce\bar{\nu_{e}}$.
$C$ represents the leading-order QCD corrections to the
$b{\rightarrow}s\gamma$ amplitude when evaluated at the $Q=m_{b}$ scale
\cite{Grinstein} \footnote{For a discussion about the uncertainties
derived from the selection of renormalization at the $Q=m_{b}$ scale
see \cite{buras}.}. In evaluating Eq. (\ref{Rat:2}) we also take
${|V_{ts}^{\ast}V_{tb}|}^{2}/|V_{cb}|^{2}=0.95\pm0.04$. Finally,
$A_{\gamma,g}$ are the coefficients of the effective operators for
the $bs\gamma$ and $bsg$ interactions; in our case, as mentioned above
we consider as relevant the contributions coming from the SM diagram
plus those with top quark and charged Higgs, and stops/scharms and
charginos running in the loop. Their expressions are given by:
\begin{eqnarray}
A^{SM}_{\gamma,g} & = &\frac{3}{2}\frac{m_{t}^{2}}{M^{2}_{W}}
f^{(1)}_{\gamma,g}\left(\frac{m^{2}_{t}}{M^{2}_{W}}\right) \nonumber \\
A_{\gamma,g}^{H^{-}}&=&\frac{1}{2}\frac{m^{2}_{t}}{m^{2}_{H}}
\left[\frac{1}{\tan^{2}\beta}f_{\gamma,g}^{(1)}\left(
\frac{m^{2}_{t}}{m^{2}_{H}}\right)+f_{\gamma,g}^{(2)}\left(
\frac{m^{2}_{t}}{m^{2}_{H}}\right)\right] \nonumber \\
A_{\gamma,g}^{\chi^{-}} &   =   & \sum_{j=1}^{2}\left\{
\frac{M^{2}_W}{M^{2}_{\chi_j}}\left[|V_{j1}|^{2}f_{\gamma,g}^{(1)}
\left(\frac{m^{2}_{\tilde{c}}}{M^{2}_{\chi_j}}\right)-
\sum_{k=1}^{2}\left|V_{j1}T_{k1}-\frac{V_{j2}m_tT_{k2}}{\sqrt{2}
M_W\sin\beta}\right|^{2} \right. \right.\nonumber \\
                        &\times & \left. \left.
f_{\gamma,g}^{(1)}\left(\frac{m^{2}_{\tilde
{t}_k}}{M^{2}_{\chi_j}}\right)\right]-\frac{U_{j2}}{\sqrt{2}\cos\beta}
\frac{M_W}{M_{\chi_j}}\left[V_{j1}f_{\gamma,g}^{3}\left(\frac
{m^{2}_{\tilde{c}}}{M^{2}_{\chi_j}}\right) \right. \right. \label{apms}	\\
                        &-      &\left.\left.\sum_{k=1}^{2}\left(
V_{j1}T_{k1}-V_{j2}T_{k2}\frac{m_t}{\sqrt{2}M_{W}\sin\beta}\right)
T_{k1}f_{\gamma,g}^{(3)}\left(\frac{m^{2}_{\tilde{t}_k}}{M^{2}_{\chi_j}}
\right)\right]\right\} \nonumber
\end{eqnarray}
where the functions $f_{\gamma,g}^{i}$, $i=1,2,3$ may be
found in \cite{Bert:Nucl}, and all the masses are understood to
be at the electroweak scale. $V$ and $U$ are the matrices which
diagonalise the chargino mass matrix, while $T$ diagonalises the
stop mass matrix.

	Our strategy is to use the renormalization group equations
(RGEs) to calculate the mass
spectrum subject to combined constraints from experiment
and  correct
radiative electroweak breaking,
using as boundary conditions for the soft-susy breaking terms
string  scenarios  obtained in ref. \cite{Iba:Spain}. We then
use this
spectrum to evaluate the ratio $R$ via Eq. $(\ref{Rat:1})$ and
Eq. $(\ref{Rat:2})$ to obtain
\begin{equation}
BR(B{\rightarrow}X_{s}\gamma)=R{\times}BR(B{\rightarrow}X_{c}e\bar{\nu_e})
\label{K:1}
\end{equation}
and compare the results with the CLEO II bound. Let us describe now
our renormalization group procedure in detail. We want of course to
integrate our RGEs from the string unification
scale $M_{string} {\rm of} O(10^{16}){\rm GeV}$ down
to the electroweak scale $M_{Z}$ \footnote{we assume that
string threshold corrections are such that they explain
the mismatch between the unification scale and the
string scale \cite{Iba:Spain}.} . In order that our numerical
integration routines make the run starting from $M_{string}$, we must
have the values of all the parameters at this scale, but this is
difficult to achieve. The problem is that the values of many
parameters are known experimentally at low scales. However,
the values of other parameters, such as soft breaking terms, are
most easily understood at higher energies where theoretical
simplification may be invoked. Thus, there is no scale at which
there is both theoretical simplicity and experimental data. In
other words, choosing $M_{string}$ as our starting point we want to
find the $M_{string}$ values of all the parameters such that we
recover the expected low energy values after renormalization
group evolution to experimental scales.
An approach that incorporates some boundary conditions at both
electroweak and $M_{string}$ scales, is the so called ambidextrous
approach \cite{Nano,Barger}. In this approach one specifies
$m_{t}$ and $\tan{\beta}$ at the electroweak scale (along with
$M_{Z}$ and $M_{W}$) and $M_{1/2}$, $m_{0}$, and $A$ at the $M_{string}$
scale. Thus first one integrates the dimensionless parameters given
$\tan{\beta}(M_{Z})$ and $m_{t}(M_{Z})$ up to $M_{string}$ in order to specify
the complete set of boundary conditions at this scale. Then all the
parameters including the soft SUSY-breaking are evolved from $M_{string}$
to the electroweak scale. At this scale $\mu(M_{Z})$ and $B(M_{Z})$
are determined by minimizing the one-loop effective potential
$V^{1-loop}$ . Subsequently $\mu$ and $B$ can be RGE-evolved up to
the string scale. This strategy is effective because the RGEs for the
soft-supersymmetry breaking parameters do not depend on $\mu$
and $B$. This method has two powerful advantages. First, any
point in the $m_{t}-\tan{\beta}$ plane can be readily investigated
in specific supergravity models since $m_{t}$ and $\tan{\beta}$ are
taken as inputs. This is extremely useful since after the recent
discovery of the top quark at Fermilab its mass is going to be
determined with high accuracy in the future. Thus approaches like
the top-down approach in which $m_t$ is output cannot effectively
scan the parameter space of supergravity models. Secondly, the
minimization conditions of $V^{1-loop}$ are easily solved for
$\mu$ and $B$. Specifically, we employ a two-dimensional Newton
method from the NAG library which quickly locates the extremal
values for $\mu$,$B$ by iteration.

\section {\bf Effective Supergravities from String Theory.}

	The low-energy limit of the supertring models relevant for the
phenomenology is the $N=1$ supergravity (SUGRA) described by the
$\rm{K\ddot{a}hler}$ function $G$, which is a function
of the $\rm{K\ddot{a}hler}$ potential $K$ and the superpotential $W$,
and the gauge
kinetic functions $f_{a}$ \cite{Cremmer:SUG}. The generic fields present
in the massless string spectrum contain the dilaton superfield $S$,
moduli fields generically denoted by $T_i$ (which can contain
the radii-type moduli $T_i$ and the complex structure moduli
$U_j$) and some matter chiral fields $\phi^{\alpha}$, containing the
Standard Model particles. The resulting effective low-energy theory,
emerging from string theory, possesses a high degree of symmetry, which
in general restricts the form of the three SUGRA functions mentioned
above. As a result, the soft parameters are also constrained \footnote
{This is to be contrasted with conventional SUGRA theories where $G$
and $f$ are arbitrary}.

	A particularly interesting class of such {\em stringy}
symmetries are the {\em target-space duality} symmetries. The physical
content of such symmetries is that in string theory physics at a very
small scale cannot be distinguished from physics at a very large scale.
Under such symmetries the moduli fields $T_i$ transform as
\begin{equation}
T_i\rightarrow\frac{a_{i}T_{i}-ib_{i}}{ic_{i}
T_{i}+d_{i}}\;\;,a_{i}d_{i}-b_{i}c_{i}=1
\;\;,a_{i}{\ldots}d_{i}{\in}Z.
\label{discre}
\end{equation}
In effective string theories of the orbifold type [13], the
matter fields $\phi^{\alpha}$ transform under (\ref{discre}) as
\begin{equation}
\phi^{\alpha}\rightarrow(ic_{i}T_{i}+d_{i})^{n_{\alpha}^{(i)}}\phi^{\alpha}
\end{equation}
where the integers $n_{\alpha}^{(i)}$ are called  modular weights (usually
are negative integers).
With the above transformations
the $\rm{K\ddot{a}hler}$ $G$ function  is modular invariant (if the
superpotential $W$ has modular weight -3).

The $\rm{K\ddot{a}hler}$ potential $K$ (to first order in the observable
fields) is given in general by the form \cite{Iba:Spain,witten}
\begin{equation}
K=-\log{(S+S^{\ast})}+K_{0}(T,T^{\ast})+K_{\alpha\beta}(T,T^{\ast})\phi_{\alpha}\phi^{\ast}_{\beta}
\end{equation}
where the indices $\alpha,\beta$ label the charged matter fields. The authors
in ref. (1) concentrated in the case of the overall modulus $T$ and
disregarded any mixing between the $S$ and $T$ fields kinetic terms
which is strictly correct at the tree level. At one loop level such
a mixing arises through the Green-Schwarz mixing coefficient
$\delta_{GS}^{i}$ \cite{Der:Gen}.

The scalar potential in the low-energy supergravity action has the
form \cite{Cremmer:SUG}
\begin{equation}
V=|W_{SUSY-breaking}(T,S)|^{2}e^{K_{0}}(G^{i}(G^{-1})_{i}^{j}G_{j}-3)
\label{dynamiko}
\end{equation}
$(e^{G}=|W|^{2}e^{K}, G_{i}=\partial{G}/\partial{\phi_{i}}.)$
In deriving (\ref{dynamiko}) the authors in \cite{Iba:Spain}
assumed that, upon minimization of
$V$,$\langle$$G_{\alpha}$$\rangle$$=0$ and
$\langle$$Q_{\alpha}$$\rangle$$=0$ in the matter sector. This
assumption, which is satisfied in most realistic scenarios,
means that the spontaneous supersymmety breaking takes place
in the dilaton-moduli sector, i.e. $\langle$$G_{i}$$\rangle$${\neq}0$
for at least one of the moduli fields.
Then the gravitino mass becomes
\begin{equation}
m_{3/2}=e^{K_0(T,S,T^{\ast},S^{\ast})/2}|W_{SUSY-breaking}(T,S)|
\end{equation}
$m_{3/2}$ should be of order TeV. Then one can obtains the following
soft terms;first the gaugino masses take the form
\begin{equation}
M_{a}(T,T^{\ast},S,S^{\ast})=
\frac{1}{2}m_{3/2}G^{i}(T,S,T^{\ast},S^{\ast})\partial_{i}\log
g_{a}^{2}(T,S,T^{\ast},S^{\ast})
\label{gaugino}
\end{equation}
The scalar masses (squarks and sleptons) become \cite{Iba:Spain}
\begin{equation}
m^{2}_{\alpha\bar{\beta}}[K_{\alpha\bar{\beta}}
(T,S,S^{\ast},T^{\ast})-G^{i}(T,S,S^{\ast},T^{\ast}
)G^{\bar{j}}(T,S,S^{\ast},T^{\ast})R_{i\bar{j}\alpha\bar{\beta}}]
\label{scalar}
\end{equation}
$(R_{i\bar{j}\alpha\bar{\beta}}=\partial_{i}{\bar{\partial}}_{\bar{j}}
K_{\alpha{\bar{\beta}}}-\Gamma^{\gamma}_{i\alpha}K_{\gamma\delta}
{\bar \Gamma}^{\bar{\delta}}_{\bar{j}\bar{\beta}}, \Gamma^{\gamma}_
{i\alpha}=K^{\gamma{\bar{\delta}}}{\partial_{i}}K_{\alpha\bar{\gamma}})$

By assuming that SUSY breaking is triggered by the auxiliary fields
of the dilaton-moduli sector, one can parametrize the unknown
supersymmetry dynamics by some angle $\tan{\theta}={\langle}F_{S}{\rangle}/
{\langle}F_{T}{\rangle}$ \cite{Iba:Spain}. Then the exact form of the
(perturbative or non-perturbative) superpotential is parametrized
by $\theta$ and $m_{3/2}$, and the form of the soft-parameters depend
only on known perturbative quantities like $K$.
Next we discuss the different scenarios that emerge in this framework
which are subject of our research in this paper.

\section {Models}

Using the general expressions (\ref{gaugino}), (\ref{scalar}) the
following form of soft terms may be derived \cite{Iba:Spain} \footnote
{we consider no-scale scenarios in which the cosmological constant is
zero. If one relaxes the constraint of the vanishing of the cosmological
constant the soft terms depend explicitly on the nonzero value of the
latter and consequently the experimental predictions of the models \cite
{Iba:Spain}.}
\begin{equation}
m_{\alpha}^{2}=m_{3/2}^{2}[1+n_{\alpha}
\cos^{2}{\theta}]
\label{vathmoto}
\end{equation}
\begin{equation}
M_{a}=\sqrt{3}m_{3/2}\frac{k_{a}ReS}{Ref_{a}}\sin{\theta}
+m_{3/2}\cos{\theta}\frac{B_{a}^{'}(T+T^{\ast})\hat{G}_{2}(T,T^{\ast})}
{32{\pi}^{3}Re{f}_{a}}
\end{equation}
\begin{equation}
A_{\alpha\beta\gamma}=-\sqrt{3}m_{3/2}\sin{\theta}-
m_{3/2}\cos{\theta}(3+n_{\alpha}+n_{\beta}+n_{\gamma})
\end{equation}
where $k_{a}$ is the Kac-Moody level of the gauge factor. In the
phenomenological analysis that follows $k_{3}=k_{2}=\frac{3}{5}k_{1}=1$
and the definitions of $B^{'}_{a}$, $\hat{G}_{2}$ functions may
be found in \cite{Iba:Spain}.

 As regards the $B$ soft term associated with the Higgs mixing
$\mu$ term in the superpotential its form is model dependent. In
particular its value depends
on the scenario we use for the
generation of the $\mu$ term \footnote {For a recent review
see Ref. \cite{ispania}}.
In string theory we have

$\bullet$ The quadratic $\mu$ term arises as an effective non-
renormalizable fourth- (or higher) order term in the superpotential
of the form
\begin{equation}
{\lambda}W_{0}H_{1}H_{2}
\end{equation}
where $W_{0}$ is the renormalizable superpotential and $\lambda$
an unknown coupling, which mixes the observable sector with the
hidden sector,
then a $\mu$ term is automatically generated
with size $\mu={\lambda}m_{3/2}$ \cite{CASAS}.

$\bullet$ The quadratic $\mu$ term is built into the theory
through the $\rm{K\ddot{a}hler}$ potential, and becomes
non-zero and of $O(m_{3/2})$ upon superymmetry breaking \cite{louis,
18,LUST}.

In no-scale scenarios the value of $B$ in both cases is given by
\begin{equation}
B=2m_{3/2}
\label{beta}
\end{equation}
As one can see the soft terms are in general ${\em
non-universal}$.
However for $\theta=\frac{\pi}{2}$, i.e the dilaton dominated
supersymmetry breaking and neglecting threshold corrections, the soft
terms are in fact universal \cite{Iba:Spain}:
\begin{equation}
m_{0}=\frac{1}{\sqrt{3}}M_{1/2},\;\;\; A=-M_{1/2}
\label{dilaton}
\end{equation}
In the strict dilaton-dominated scenario the $B$ soft term is
also predicted
to have the value
\begin{equation}
B=2m_{0}=\frac{2}{\sqrt{3}}M_{1/2}
\end{equation}
However, in the general dilaton-dominated scenario the $B$ term is
an independent parameter.
The latter scenario is subject of our research
in this paper. In the numerical approach we use  the $B$
term is determined by the minimization conditions.

The special of properties of the dilaton dominated scenario have been
recently {\em emphasized} in ref. \cite{Ib:Madr}. It is also worthwhile
to reiterate that the dilaton dominated scenario is of general validity
since the boundary conditions in (\ref{dilaton}) are obtained for
{\em any} 4-D N=1 string and not only for orbifolds. An initial study
of the phenomenology of the above soft terms  in the context of the
Minimal Supersymmetric Standard Model (MSSM)
 has been done in \cite{Bar:Pisa,Iba:Spain}. In ref.\cite{LN:austin},
Lopez et al, studied the phenomenological consequences of (\ref{dilaton})
in the context of $SU(5)\times U(1)$ which predicts extra matter particles
\footnote {In string models derived in the fermionic formulation
the extra matter hypothesis is imperative for reconciling the string
unification scale with the LEP data, at least at the $k=1$ level
\cite{Farragi}.}.

\section{Analysis}
As was said in the introduction we use the ambidextrous approach in our
RG analysis. Thus, our parameter space in the dilaton-dominance limit
is $\tan{\beta}$, $m_{t}(M_{Z})$,  $M_{1/2}$
and the sign of $\mu$ which is not determined by the radiative
electroweak breaking constraint. In the more general
case where the moduli also contribute to SUSY breaking the goldstino
angle is added to the parameter space.
In the latter case, we must take into account additional $D-$ term
contributions to the scalar masses due to the non-universality of the
scalar soft-terms in (\ref{vathmoto}) \cite{Munoz,Kelley}. In
particular, the combination
\begin{eqnarray}
{\cal S} &=&m^{2}_{H_{2}}-m^{2}_{H_{1}} \nonumber \\
  &+&\rm {Tr}[{\bf M}^{2}_{Q_{L}}-{\bf M}^{2}_{L_{L}}-2{\bf
M}^{2}_{U_{R}}+{\bf M}^{2}_{D_{R}}+{\bf M}^{2}_{E_{R}}]
\end{eqnarray}
contributes to the RGEs and satisfies the (one-loop) scaling equation
\begin{equation}
\frac{d{\cal S}}{dt}=\frac{2b_{1}g_{1}^{2}}{16{\pi}^{2}}{\cal S}
\label{nonuni}
\end{equation}
so that if it is zero at some scale, for example the string scale,
then it is zero for all scales. The renormalization group coefficient
$b_{1}=33/5$ in the MSSM. In the dilaton-dominated scenario the
${\cal S}$ term does not contribute to the scaling of scalar masses. In
the non-universal case we consider the model with modular weights
$n_{\alpha}$ different from -1 which was first
studied in \cite{Iba:Spain} and gives unification at a scale of
$O(10^{16})GeV$. Again we prefer to allow the $B$ soft term to
be a free parameter
given the uncertainty conserning
the $\mu$ term. However, the results as regards
the $b{\rightarrow}s\gamma$ decay in the latter model are similar
to those obtained in the dilaton-dominated scenario. The reason
is that the $\sin{\theta}$ parameter takes values close to one
in order to avoid tachyonic states
and therefore the dilaton $F-$
term is the dominant source of supersymmetry
breaking see Brignole et al in \cite{Iba:Spain}.

The string low energy observable sector is identified with that
of the MSSM, and the perturbative superpotential which describes
the renormalizable trilinear and bilinear Yukawa couplings of
quarks,leptons and Higgs bosons chiral superfields
is given by
\begin{eqnarray}
W
&=&\sum_{i}{h_{u_{i}}Q_{i}H_{2}U^{c}_{i}+h_{d_{i}}Q_{i}H_{1}D^{c}_{i}+h_{e_{i}}L_{i}H_{i}E^{c}_{i}} \nonumber \\
  &+& {\mu}H_{1}H_{2},
\label{super:1}
\end{eqnarray}
where $i$ is a generation index, $Q_{i}(L_{i}$ are the scalar partners
of the quark (lepton) SU(2) doublets, $U_{i}^{c},D_{i}^{c}(E_{i}^{c})$
are the quark (lepton) singlets and $H_{1,2}$ are the two supersymmetric
Higgs doublets. The $h-$ factors are the Yukawa couplings and $\mu$
is the usual Higgs mixing parameter. In Eq. $(\ref{super:1})$ the usual
SU(2) contraction is assumed, e.g $\mu\epsilon_ijH^{i}_{1}H^{j}_{2}$
with $\epsilon_{12}=-\epsilon_{21}=1$. Then the Lagrangian will contain
besides the superymmetric $F-$ and $D-$ terms the following soft
supersymmetric breaking terms
\begin{eqnarray}
{\cal L}_{SB}&=& \frac{1}{2}M_a{\lambda_{a}}\lambda_{a}+
\left\{\sum_{i,j}(m^{2})^{j}_{i}\phi^{i}\phi_{j} \right.\nonumber \\
            & + &\left.\sum_{i}[A_{u_i}h_{u_i}{\tilde Q}_{i}H_{2}{\tilde
U}^{c}_{i}+
A_{d_i}h_{d_{i}}{\tilde Q}_{i}H_{1}{\tilde D}^{c}_{i}+
A_{e_{i}}h_{e_{i}}{\tilde L}_{i}H_{1}{\tilde E}^{c}_{i}+ {\rm {h.c}}] \right.
\nonumber \\
           &  + &
\left.\left[B{\mu}H_{1}H_{2} +{\rm {h.c}}\right]\right\}
\;\;,
\end{eqnarray}
where $\phi_{i}$ denotes a generic scalar field.

For the study of radiative electroweak breaking constraint (REWB) we use
the one-loop effective potential instead of the tree-level potential
$V_{0}$
\begin{equation}
V^{1-loop}=V_{0}+\Delta{V_{1}}
\end{equation}
where
\begin{equation}
\Delta{V_{1}}=\frac{1}{64{\pi^{2}}}{\rm Str}\left[{\cal M}^{4}
\left(\log{\frac{{\cal M}^{2}}{Q^{2}}}-\frac{3}{2}\right)\right]
\label{correction}
\end{equation}
depend on the Higgs fields through the tree-level squared-mass
matrix ${\cal M}^{2}$. The supertrace in (\ref{correction})
is given by
\begin{equation}
{\rm Str}f({\cal M}^{2})=\sum_{i}(-1)^{2J_{i}}(2J_{i}+1)f(m_{i}^{2})
\end{equation}
where $m_{i}^{2}$ denotes the field-dependent mass eigenvalue of
the $\it{i}$th particle of spin $J_{i}$. For the calculation of
radiative corrections we use the tadpole method \cite{Weinberg} which
is a very convienient way of incorporating the corrections into
the minimization conditions of all the particle spectrum.

The chargino mass term in matrix form, which plays a crusial role
in the expressions for $BR(B{\rightarrow}s\gamma)$ [see Eq.(3)]
is given by
\begin{center}
$\left(
\begin{array}{cc}
M_{2}                        &     \sqrt{2}m_{W}\sin{\beta} \\
\sqrt{2}m_{W}\cos{\beta}     &        -\mu
\end{array}
\right) $
\end{center}

Besides the constraint of correct electroweak breaking,
the experimental constraints we impose in the above superstring
scenarios are
(1) We require that all sleptons be heavier than $M_{Z}/2$, since
sleptons are not observed in Z decays \cite{ALEPH}.
(2) We require that the lightest chargino mass eigenstate,
$M_{{\tilde \chi}_{1}^{+}}$, be heavier than $M_{Z}/2$, since
chargino pairs are not observed in Z decays \cite{ALEPH}.
(3) We impose that gluinos be heavier than 120GeV. However, this
requirement is not so constraining since the sleptons and chargino
boundary conditions require that $M_{g}$$>$$200$ GeV. Because
of naturalness criteria the largest gluino masses we study
correspond to $M_{g}{\approx}\rm {1TeV}$.
(4) As regards the Higgs sector we require that the lightest Higgs
eigenstate, $h^{0}$, is heavier than 60GeV and that the CP-Odd mass
eigenstate $A^{0}$ is not visible at LEP.
(5) We demand that all squarks should be heavier than 45GeV
and the lighest neutralino be heavier than 20 GeV, and $m_{top}=178GeV$
(6) Finally, as was said in the introduction we impose the
current CLEO bounds on the $BR(b{\rightarrow}s\gamma)$

As one can see from the graphs we plotted the values
of $BR(b{\rightarrow}s\gamma)$ vs the gluino mass
,in the dilaton-dominated scenario
and in the non-universal case with $\theta=\frac{2\pi}{3}$,for
selected values
of $\tan{\beta}$ and for the top mass $m_{t}^{\rm pole}=178GeV$
consistent with the experimental values that were announced from CDF
recently \cite{disney}. The SM prediction for
the $BR$ and the CLEO bounds are also
shown . From the graphs is evident that the CLEO upper and lower
bounds restrict the allowed parameter space dramatically and in fact
require $\mu$ to be negative. For $\mu$$>0$
the values of $BR(b{\rightarrow}s\gamma)$ increase steadily with
$\tan{\beta}$ and fall outside the experimentally allowed region
for all values of $M_g$ and for $\tan{\beta}\geq{2}$. Thus, we
see that the upper CLEO bound together with the recently announced
value for the top quark mass \cite{disney} exclude the positive branch of
the Higgs mixing parameter $\mu$.
For $\mu$$<0$ the $\tan{\beta}-$dependence is different. One sees
that $BR(b{\rightarrow}s\gamma)$ can be suppressed much below
the lower CLEO bound and consequently of the Standard Model result.
This phenomenon has been explained in \cite{Caristo,Vissani}. The
chargino contribution to the amplitude in Eq. (\ref{apms}),
can have the same sign (negative) or opposite
sign (positive) compared to $t-W^{\pm}$ and $t-H^{+}$ contributions
which are always negative. Actually, the region in which the
chargino amplitude gives rise to a destructive interference effect
with the other amplitudes corresponds to the region in which $\mu$
is negative \footnote {Notice that the renormalization for $\mu$ is
in fact multiplicative.} Thus, constructive interference occurs
for $\mu$$>0$ and destructive interference occurs for $\mu$$<0$, as
evident from the figures.

Furthermore, for the phenomenologically prefered negative branch
of the $\mu$ Higgs mixing term, we observe a tendency towards smaller
values of the ratio of the two vacuum expectation values. As
$\tan{\beta}$ increases a larger portion of the parameter space is
excluded and higher gluino masses are
preferred. Actually ${tan{\beta}}_{MAX}{\approx}30$ in this case
since higher values of this parameter which give correct electroweak
breaking are very expencive and demand very high gluino masses, above
the naturaleness bound of 1TeV.
Here, the lower CLEO bound is relevant for the constraints
described.
Thus, we can conclude that the experimental evidence for the inclusive
$b{\rightarrow}s\gamma$ decay together with the recent top quark
discovery remain among the most relevant tests for exploring
the parameter space of superstring scenarios.

\section{Acknowledgements}
I wish to thank Professor D. Bailin for many useful comments and
for reading the manuscript
and Dr Tom Browne of Computing Service, University of Sussex for
his invaluable help with the graphs.

\newpage
\begin{figure}
\vspace*{15cm}
\hspace*{0.01cm} \special {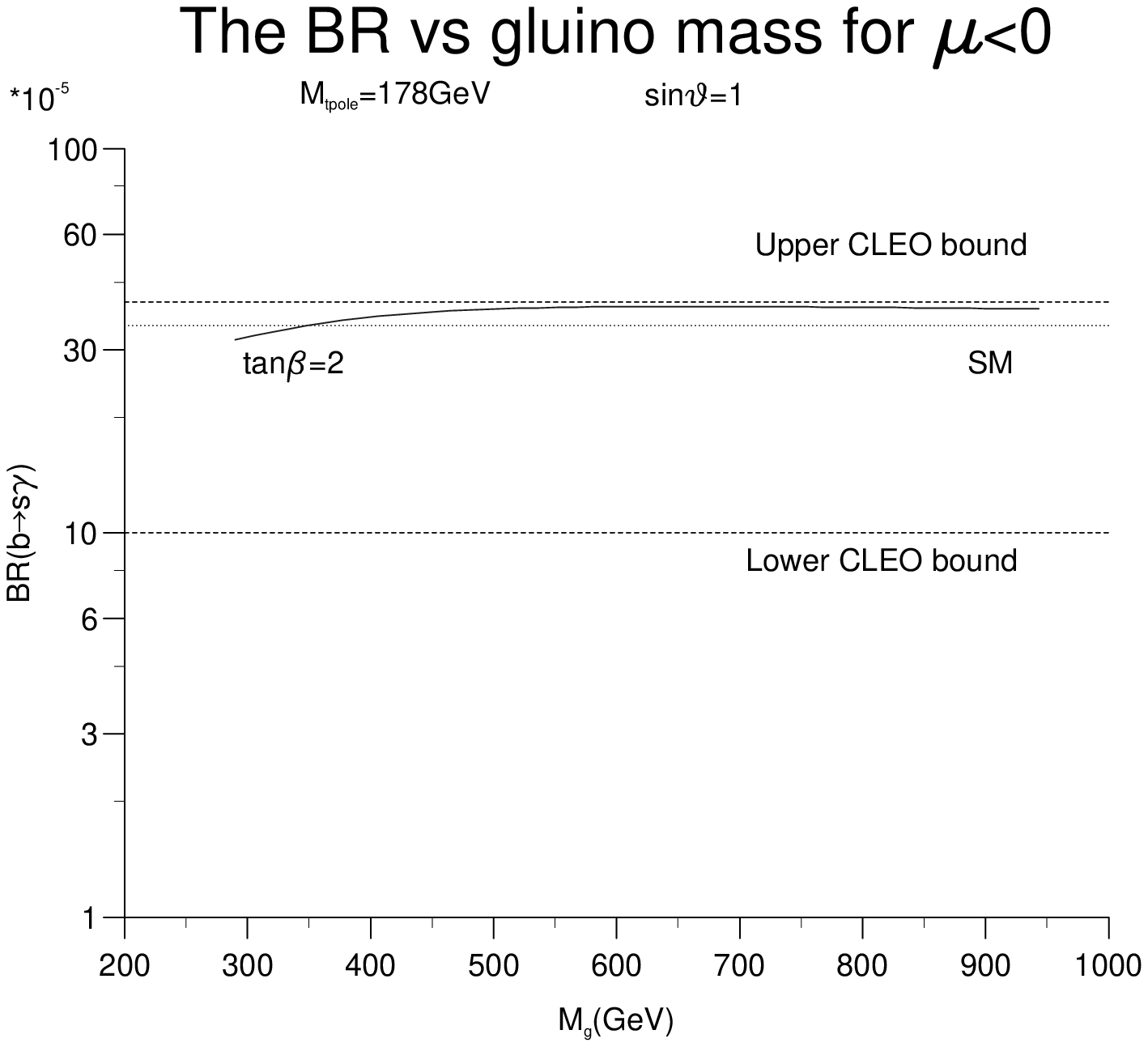}
\vspace*{1cm}
\caption{The BR in the dilaton-dominated scenario for $\tan{\beta}=2$
and $\mu$$<0$}
\end{figure}
\newpage
\begin{figure}
\vspace*{15cm}
\hspace*{0.01cm} \special {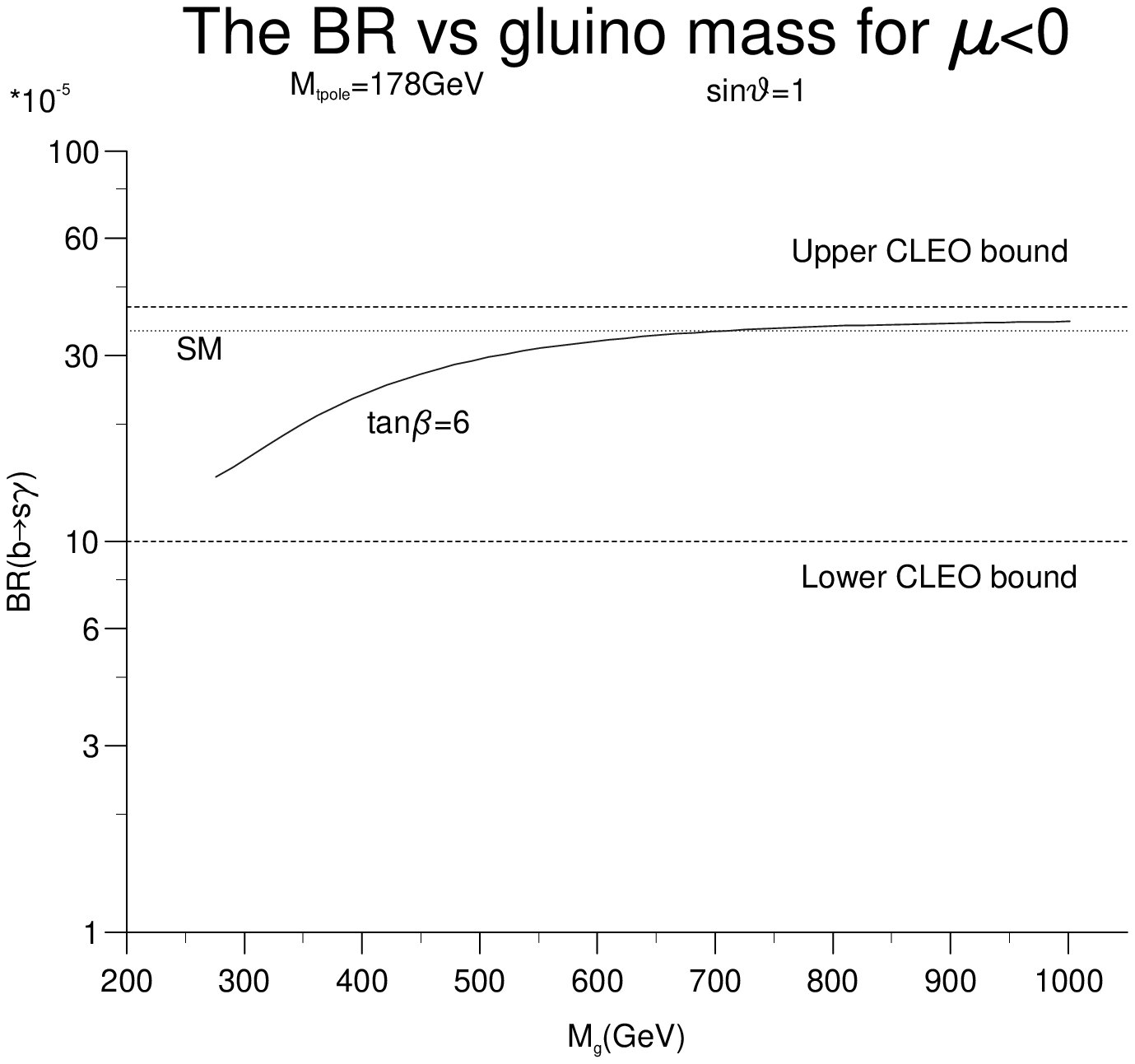}
\vspace*{1cm}
\caption{The BR in the dilaton-dominated scenario for $\tan{\beta}=6$
and $\mu$$<0$}
\end{figure}
\newpage

\begin{figure}
\vspace*{15cm}
\hspace*{0.1cm} \special {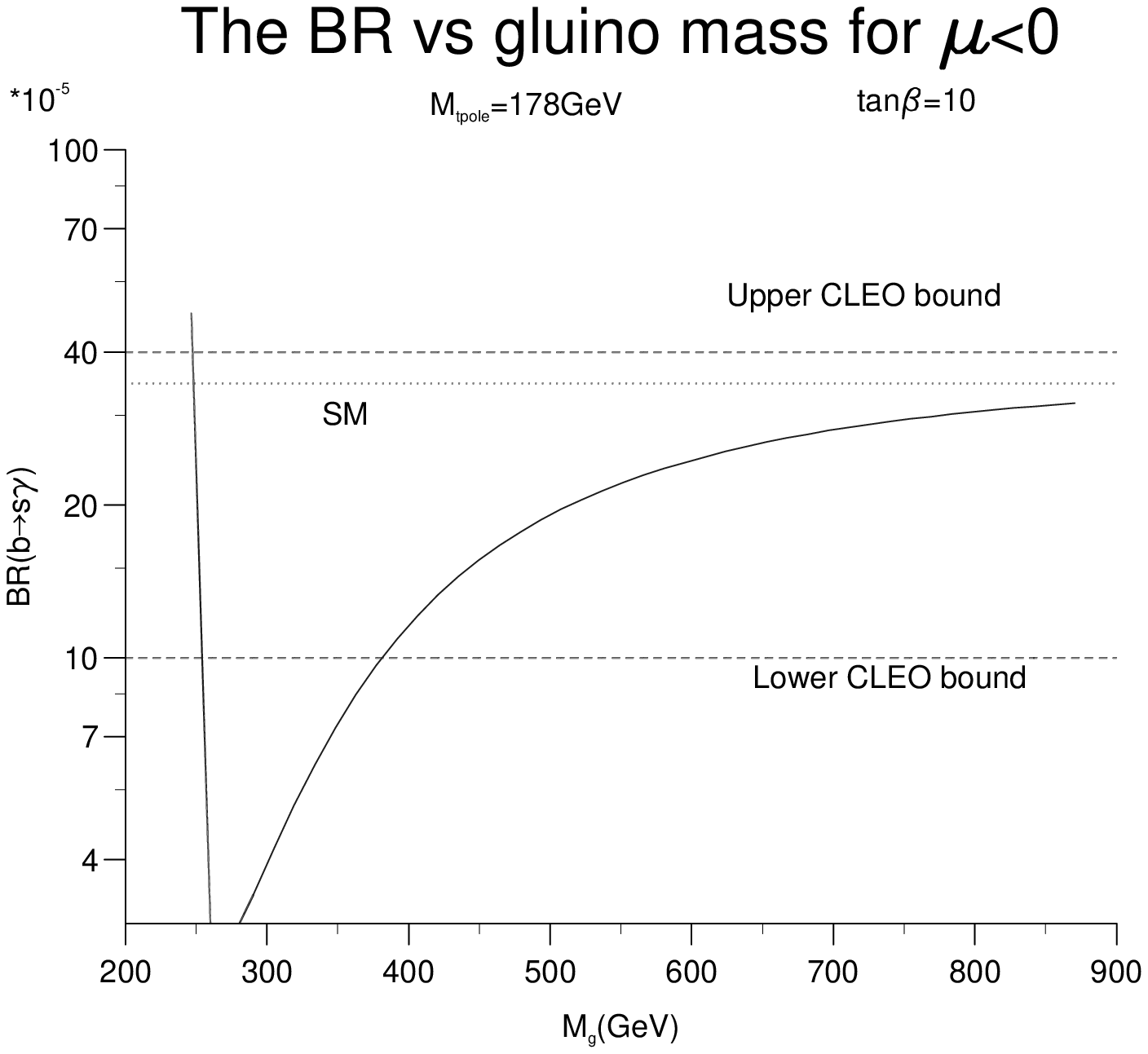}
\vspace*{1cm}
\caption{\em The ratio in the dilaton-dominated scenario for $\mu$$<0$,
$\tan{\beta}=10$}
\end{figure}
\newpage
\begin{figure}
\vspace*{15cm}
\hspace*{0.01cm} \special {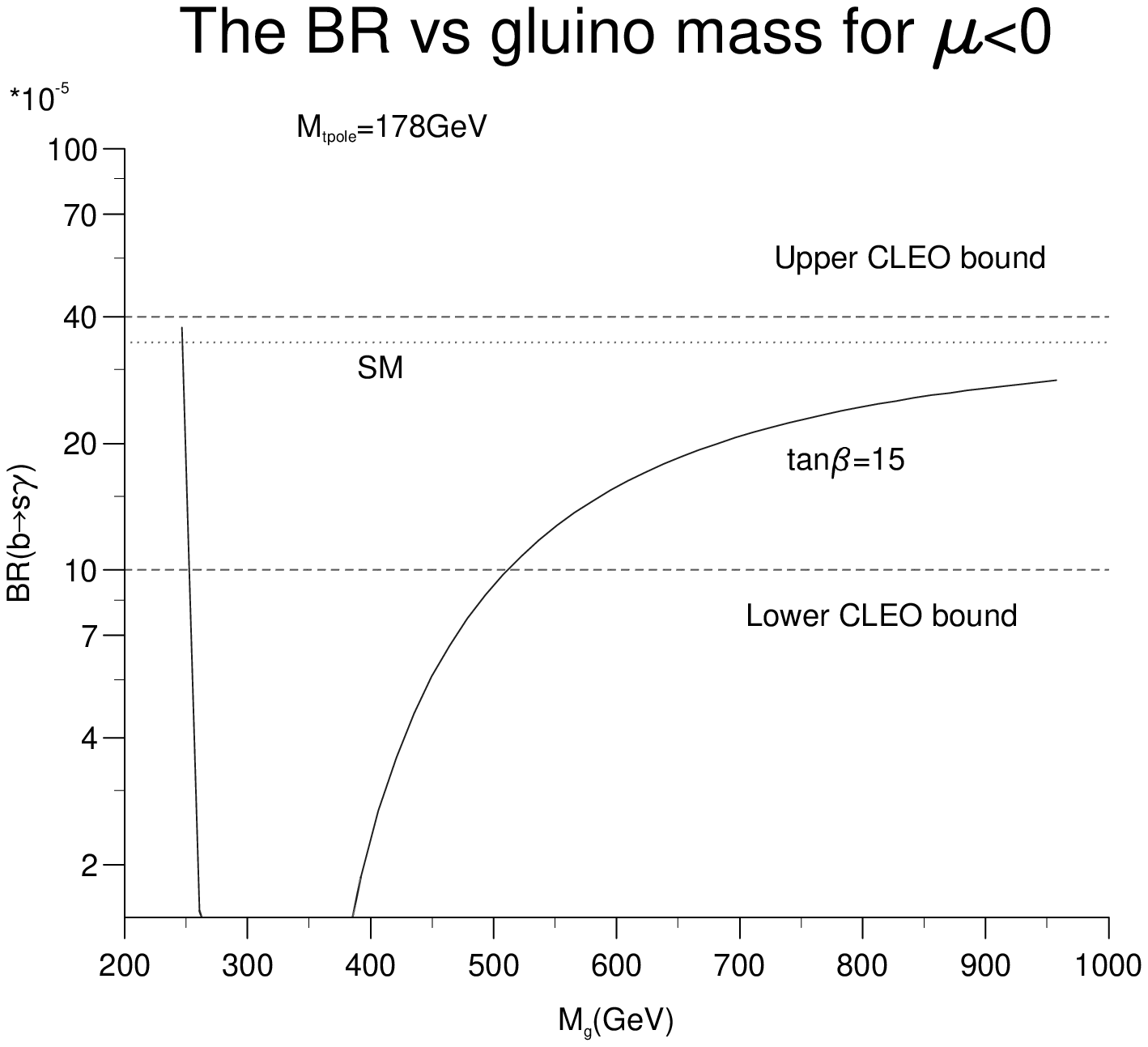}
\vspace*{1cm}
\caption{The BR in the dilaton-dominated scenario for $\tan{\beta}=15$,
$M_{tpole}=178$GeV,and $\mu$$<0$}
\end{figure}
\newpage
\begin{figure}
\vspace*{15cm}
\hspace*{0.01cm} \special {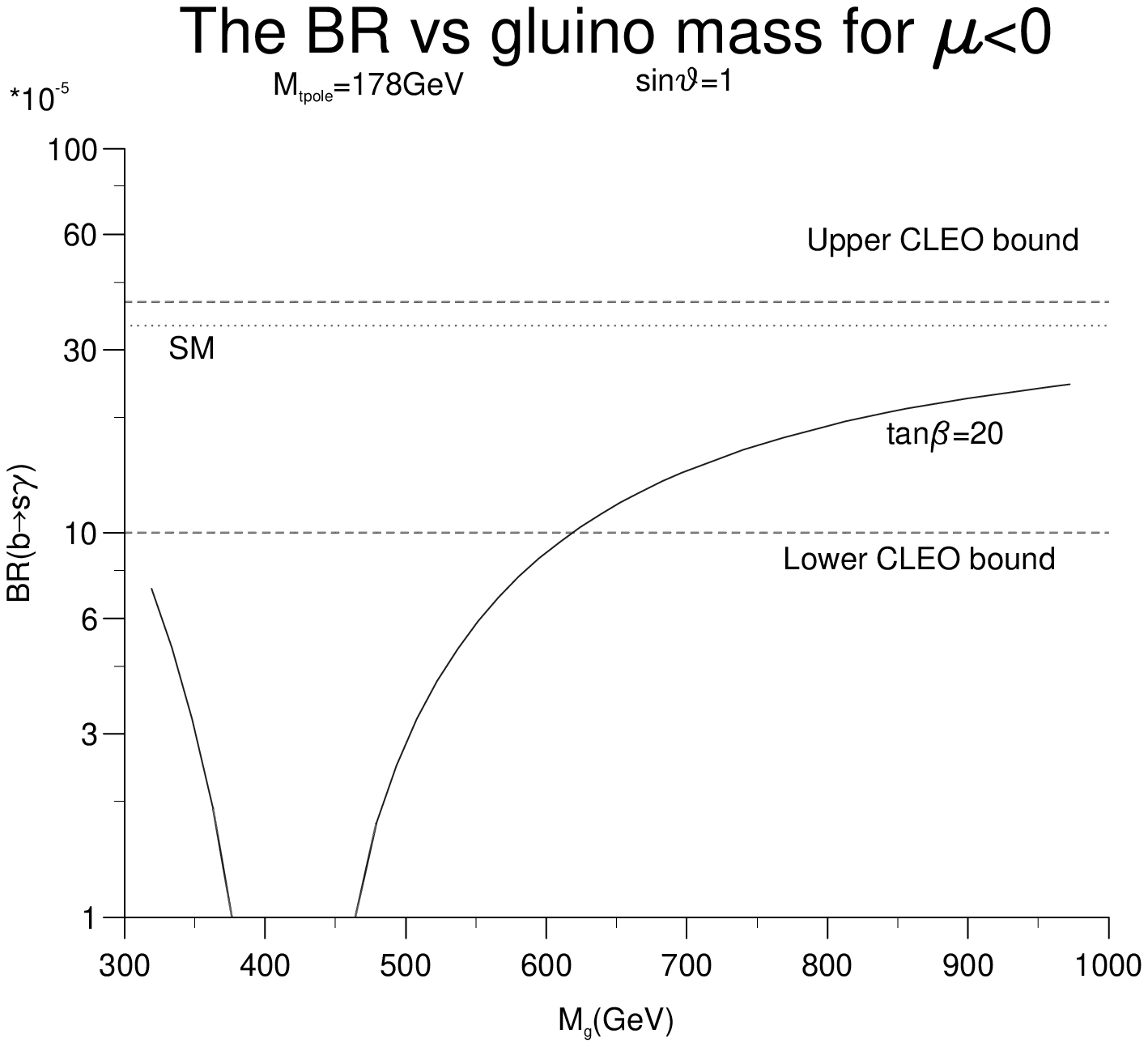}
\vspace*{1cm}
\caption{The BR in the dilaton-dominated scenario for $\tan{\beta}=20$
and $\mu$$<0$}
\end{figure}

\newpage
\begin{figure}
\vspace*{15cm}
\hspace*{0.01cm} \special {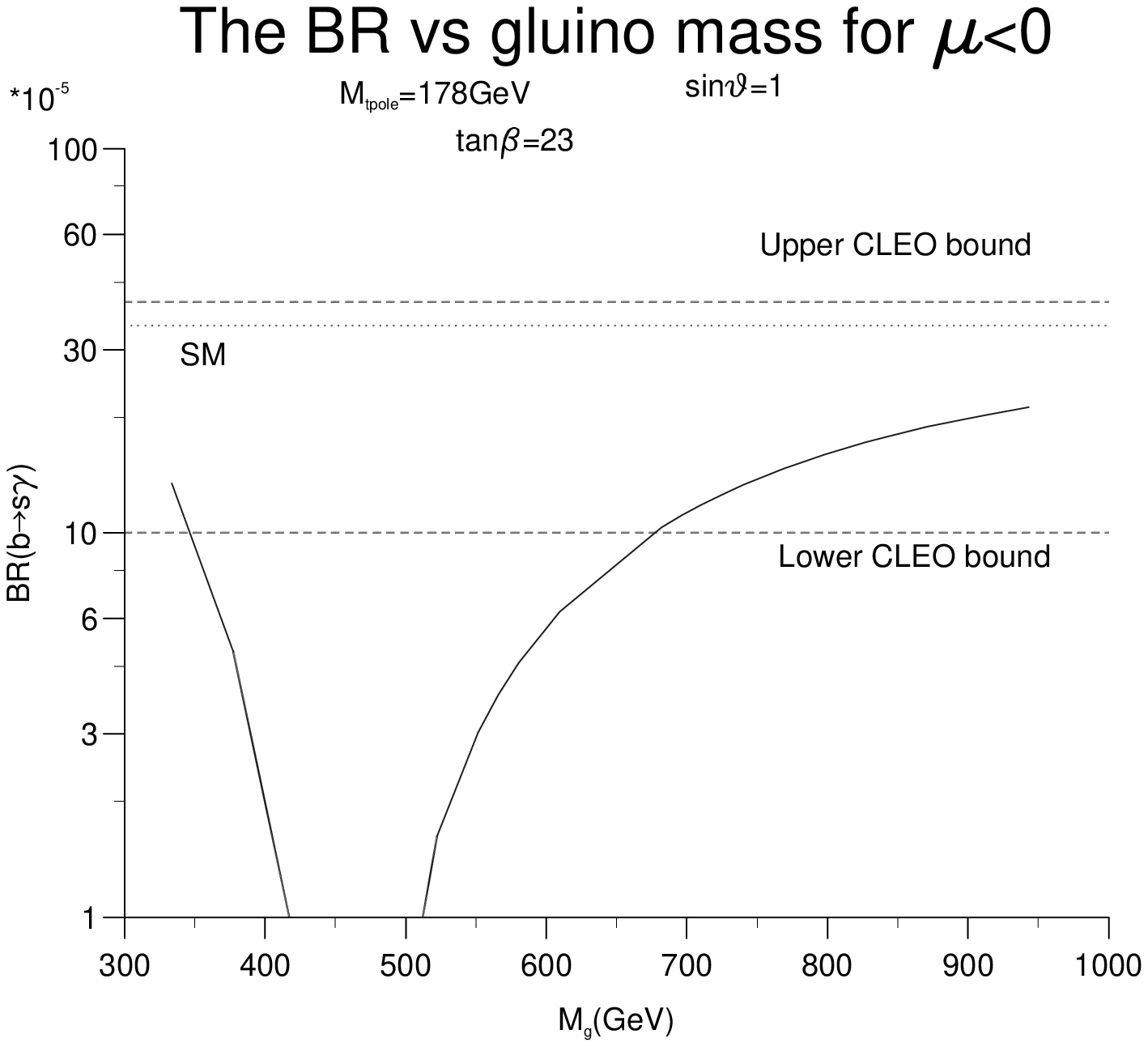}
\vspace*{1cm}
\caption{The BR in the dilaton-dominated scenario for $\tan{\beta}=23$
and $\mu$$<0$}
\end{figure}
\newpage
\begin{figure}
\vspace*{15cm}
\hspace*{0.1cm} \special {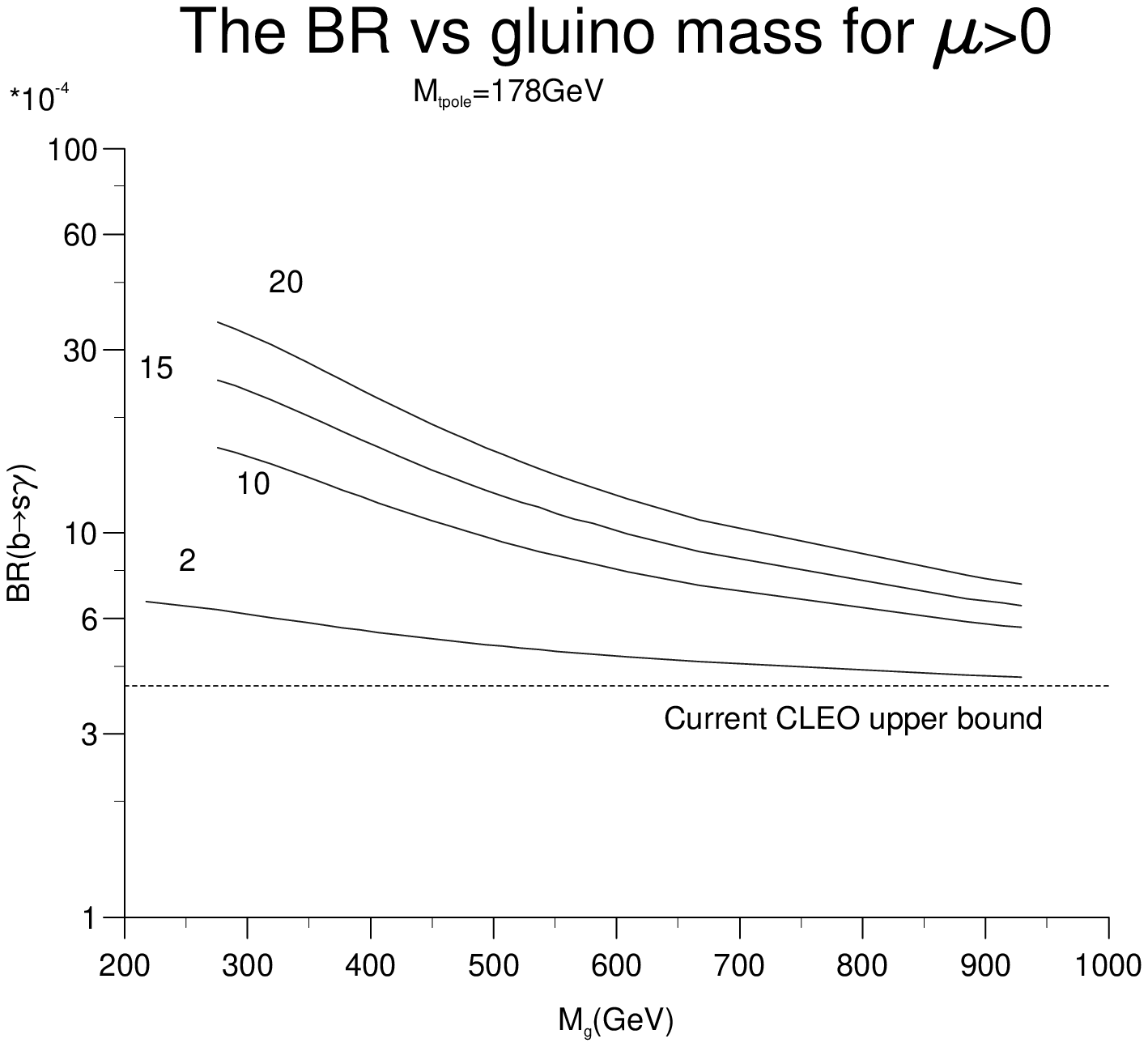}
\vspace*{1cm}
\caption{\it The ratio in the dilaton-dominated scenario for $\mu$$>0$
and different choices for
$\tan{\beta}$  ${|V^{\ast}_{ts}V_{tb}|}^{2}/|V_{cb}|^{2}=0.99$.}
\end{figure}
\newpage
\begin{figure}
\vspace*{15cm}
\hspace*{0.01cm} \special {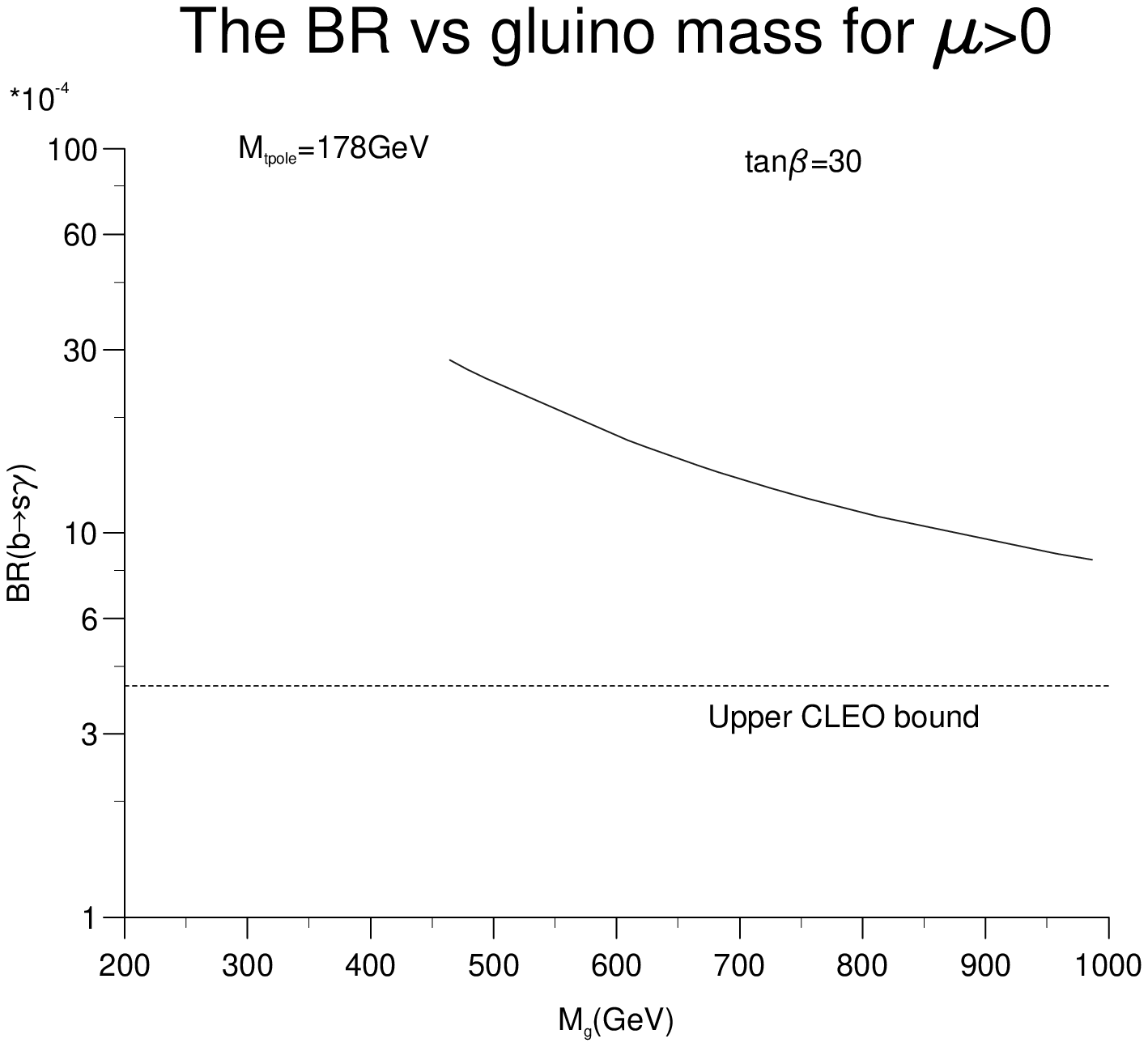}
\vspace*{1cm}
\caption{\it The ratio in the dilaton-dominated scenario for
$\tan{\beta}=30$,$\mu$$>$$0$}
\end{figure}
\newpage
\begin{figure}
\vspace*{15cm}
\hspace*{0.01cm} \special {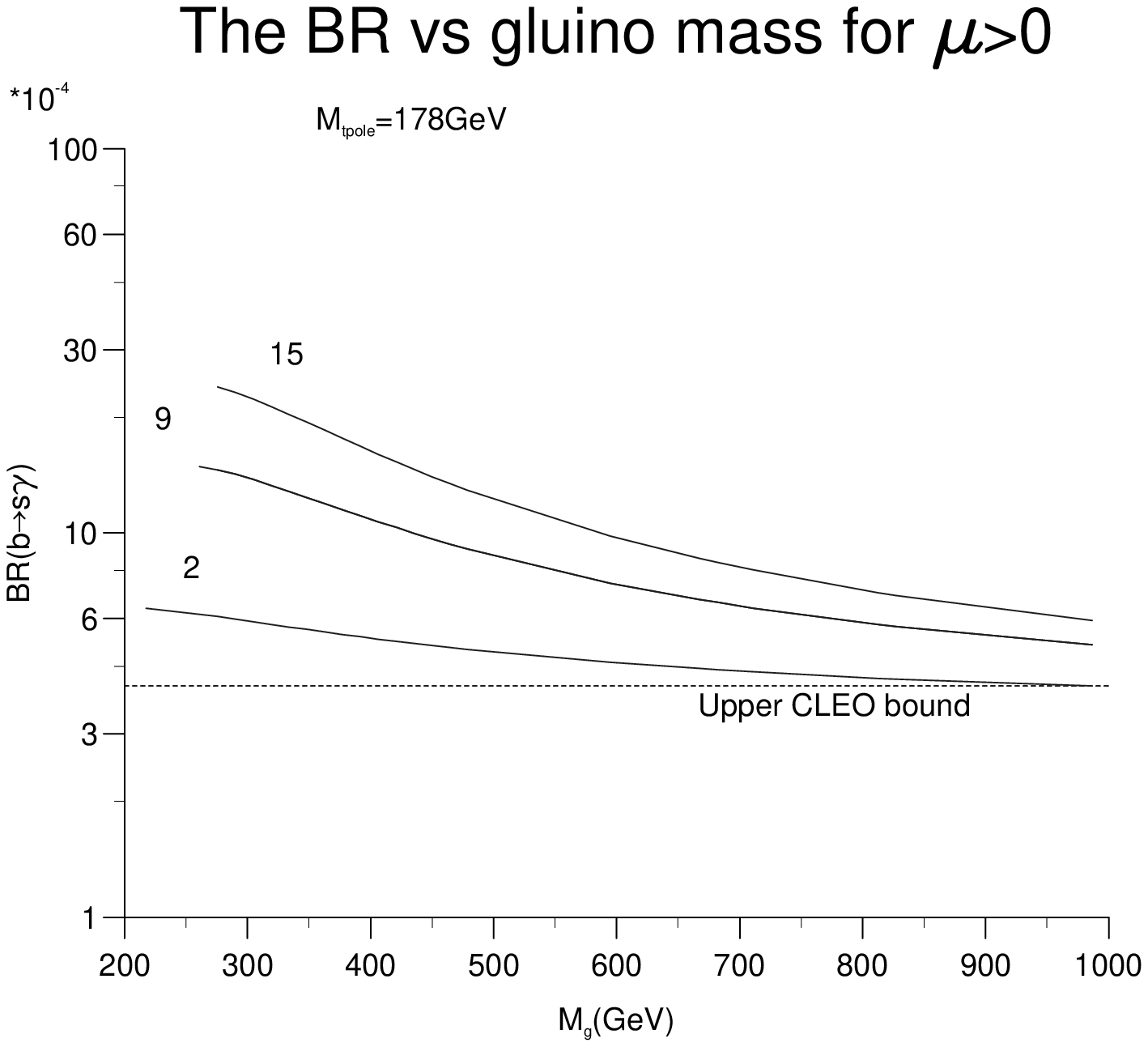}
\vspace*{1cm}
\caption{\it The ratio in the dilaton-dominated scenario for
$\tan{\beta}=2,9,15$ and ${|V^{\ast}_{ts}V_{tb}|}^{2}/|V_{cb}|^{2}=0.95$}
\end{figure}

\newpage
\begin{figure}
\vspace*{15cm}
\hspace*{0.01cm} \special {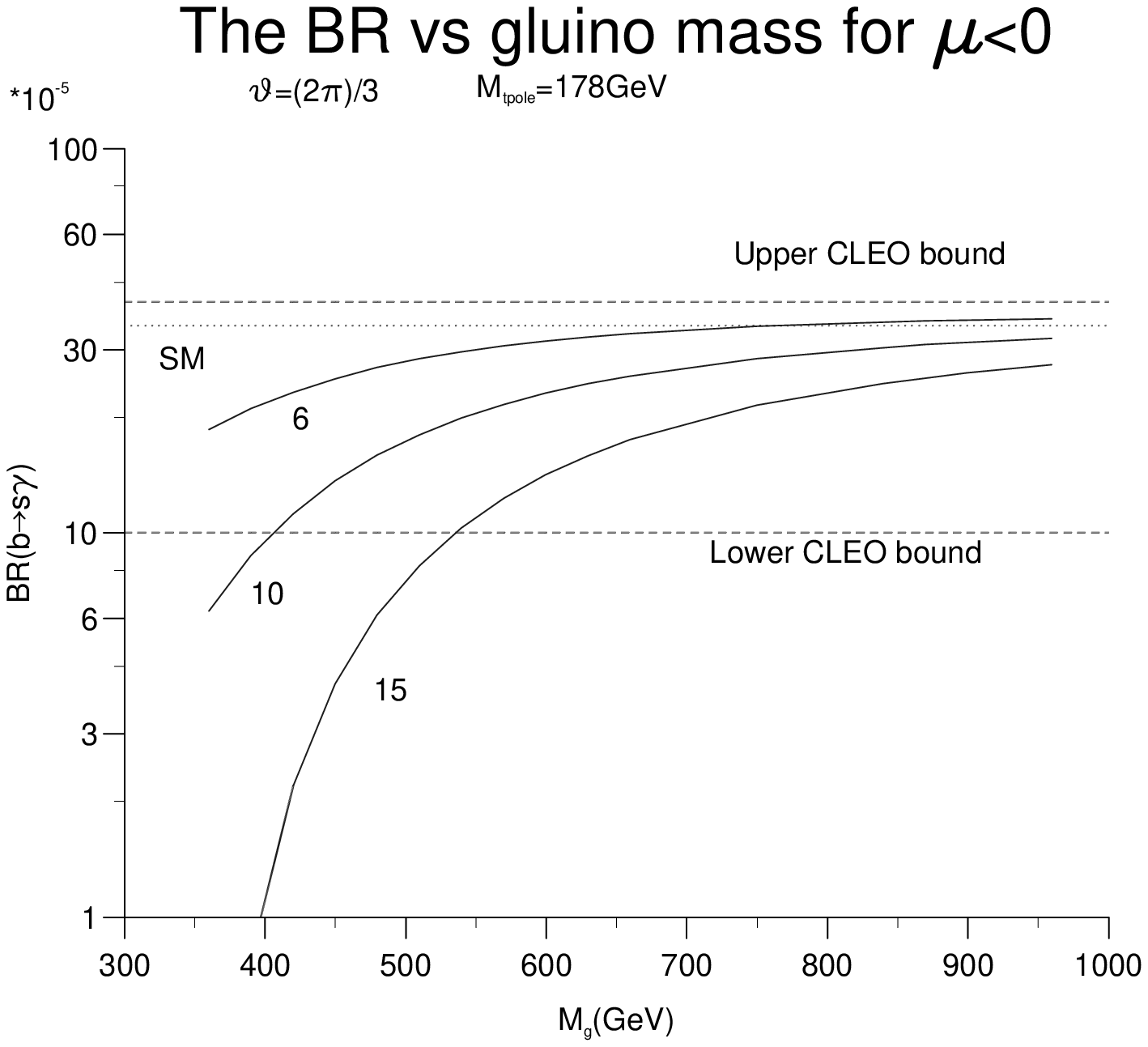}
\vspace*{1cm}
\caption{The BR in the mixed scenario, $\theta=\frac{2\pi}{3}$,
$\mu$$<0$ and different choices of $\tan{\beta}$}

\end{figure}

\end{document}